\documentclass[journal]{IEEEtran}
\ifCLASSINFOpdf
\else
\fi
\usepackage[utf8]{inputenc}
\usepackage{amsmath,amssymb,amsfonts}
\usepackage{cite}

\usepackage{algorithmic}
\usepackage{stfloats}
\usepackage{graphicx}
\usepackage{mathabx}
\usepackage{textcomp}
\usepackage{xcolor}
\usepackage{color}
\usepackage{pifont}
\usepackage{amsthm}

\usepackage{hyperref}
\usepackage{mathrsfs}
\usepackage{booktabs}
\usepackage{hyperref}
\usepackage{cases}
\usepackage{comment}
\usepackage{empheq} 
\usepackage{caption}
\usepackage{subcaption}
\usepackage{extarrows}
\allowdisplaybreaks[4]
\usepackage{url}  
\usepackage{graphicx}  
\usepackage[ruled,vlined,linesnumbered]{algorithm2e}

\begin{document}
\captionsetup[figure]{labelfont={rm},labelformat={default},labelsep=period,name={Fig.}}
\title{Mixture-of-Experts Deep Reinforcement Learning for Reliability-Constrained Energy-Efficient PDCCH Monitoring in Internet of Thing Device}

\author{Yue,Xiu~\IEEEmembership{Member,~IEEE}, Ning Wei,~\IEEEmembership{Member,~IEEE},  Meng Li, Chao Fan, Chao Wang,\\ Haopei Deng, Jinghua Zhang, Kun Xiao\vspace{-3mm}\\
\thanks{Ning Wei, Yue Xiu and Kun Xiao are with 
National Key Laboratory of Science and Technology on Communications, University of Electronic Science and Technology of China, Chengdu 611731, China (e-mail:  xiuyue12345678@163.com,
wn@uestc.edu.cn, xiaokun@uestc.edu.cn).
Meng Li, Chao Fan and Chao Wang are with 
China Mobile Internet of Things Co., Ltd. (e-mail: limeng@cmiot.chinamobile.com,
fanchao@cmiot.chinamobile.com, wangchao@cmiot.chinamobile.comn).
Haopei Deng and Jinghua Zhang are with 
Xinyi Information Technology (Shanghai) Co., Ltd. (e-mail: denghp@xinyisemi.com, zhangjh@xinyisemi.com).

}
}

\maketitle

\begin{abstract}
The continuous monitoring of the physical downlink control channel (PDCCH) is a major source of energy consumption in fifth-generation (5G) internet of thing device (IoT-D), since the UE has to blindly detect downlink control information even when no valid scheduling grant is present. Although predictive dynamic power management can reduce unnecessary receiver activity by skipping PDCCH monitoring in grant-free slots, aggressive sleeping may lead to missed grants and degrade reception reliability. To address this tradeoff, this paper formulates UE-side PDCCH monitoring as a reliability-constrained long-term energy minimization problem. Specifically, the IoT-D determines, before observing the actual scheduling outcome, whether to monitor the PDCCH or switch the receiver chain into a low-power state. The objective is to minimize the long-term average energy consumption, including receiver operating energy, component switching energy, and prediction-related computational energy, while ensuring that the false negative rate of scheduling-grant detection remains below a prescribed threshold. The resulting problem is non-convex due to the bursty and temporally correlated nature of grant arrivals, and the binary monitoring decisions coupled by a long-term reliability constraint. To solve this problem, we propose a mixture-of-experts input-output hidden Markov model (MoE-IOHMM)-based predictive monitoring scheme, where multiple IO-HMM experts capture heterogeneous grant-arrival patterns and a gating network adaptively combines their predictions.  Simulation results show that the proposed scheme effectively reduces IoT-D-side energy consumption compared with always-on PDCCH monitoring and conventional predictive baselines, while maintaining the false negative rate below the prescribed reliability threshold.
\end{abstract}

\begin{IEEEkeywords}
PDCCH, IoT-D, 5G, MoE, IOHMM
\end{IEEEkeywords}

\section{Introduction}
With the rapid deployment of fifth-generation (5G) New Radio (NR) systems, Internet of thing device (IoT-D) energy consumption has become an increasingly critical issue for mobile terminals, especially for battery-powered smartphones and Internet-of-Things devices\cite{b01}. A significant portion of IoT-D energy is consumed by the cellular modem, where the receiver chain must be frequently activated to monitor control information. In 5G NR, before receiving downlink or uplink scheduling information, the IoT-D needs to blindly monitor the physical downlink control channel (PDCCH) to detect downlink control information (DCI)\cite{b02,b03}. Since the IoT-D does not know in advance whether a valid scheduling grant will be assigned in the current transmission time interval (TTI), it has to keep the radio-frequency receiver and baseband receiver active for PDCCH monitoring\cite{b04,b05}. Even when no valid grant is present, this blind detection procedure still consumes non-negligible energy, resulting in inefficient receiver operation and reduced battery lifetime. 

Dynamic power management (DPM) provides a promising solution to reduce unnecessary modem energy consumption by switching receiver components into low-power states when communication activity is unlikely to occur\cite{b06}. Compared with passive DPM, which turns off receiver components only after confirming that no useful control or data information exists, predictive DPM attempts to make an early decision before the actual scheduling outcome is observed. If the IoT-D can predict that no valid scheduling grant will appear in the upcoming TTI, it may skip PDCCH monitoring and place the receiver chain into a low-power mode earlier\cite{b07}. In this way, the IoT-D can reduce the operating time of power-hungry components such as RF-RX and PHY-RX. However, this energy-saving opportunity comes with an inherent reliability risk. If the IoT-D skips PDCCH monitoring while a valid grant is actually present, the IoT-D will miss the scheduling opportunity, leading to a false negative event. Therefore, IoT-D-side power saving cannot be treated as a pure energy minimization problem; it must explicitly account for the missed-grant risk\cite{b08}.

Despite these useful modeling and algorithmic efforts, a clear optimization formulation is still needed to explicitly characterize the fundamental tradeoff between IoT-D energy saving and grant-detection reliability\cite{b09}. Prediction accuracy alone is not an appropriate optimization objective, especially because valid scheduling grants may be sparse in practical traffic traces\cite{b010}. A predictor that aggressively classifies most TTIs as grant-free may achieve low energy consumption but can result in an unacceptable FNR. Conversely, always monitoring the PDCCH eliminates missed grants but fails to provide any energy-saving gain. Therefore, the key question is not merely how to predict grant arrivals, but how to determine the PDCCH monitoring action under a reliability constraint\cite{b011}.

Motivated by the above studies, this paper formulates IoT-D-side PDCCH monitoring as a reliability-constrained long-term energy minimization problem. In each TTI, the IoT-D determines whether to monitor the PDCCH or switch the receiver chain into a low-power state before the actual grant state is observed. The objective is to minimize the long-term average IoT-D-side energy consumption, including receiver operating energy, component switching energy, and prediction-related computational energy. Meanwhile, the long-term FNR is constrained below a prescribed reliability threshold, ensuring that energy saving does not come at the cost of excessive missed grants. This formulation directly connects predictive grant modeling with modem-level power control and provides a principled basis for evaluating energy-efficient PDCCH monitoring policies.

The resulting problem is challenging for several reasons. First, the actual scheduling grant state is unavailable when the monitoring decision is made. The IoT-D can only rely on historical observations such as previous grant outcomes, PDCCH monitoring results, scheduling-request-related information, and other accessible modem-side features. Second, the scheduling process is controlled by the gNB-side MAC scheduler, whose internal state depends on traffic demand, channel condition, QoS requirements, retransmission status, and multi-user competition. These factors are only partially observable at the IoT-D side. Third, grant arrivals are temporally correlated and bursty rather than independent across TTIs, which makes memory-aware sequential decision making necessary. Finally, the monitoring action is binary, while the FNR constraint couples decisions across time, leading to a constrained partially observable sequential optimization problem. These challenges indicate that IoT-D-side PDCCH monitoring
cannot be effectively addressed by a purely prediction-oriented
method or a fixed sleeping rule. Instead, the monitoring policy
should jointly account for the uncertainty of grant arrivals,
the temporal dependence of scheduling behavior, and the
long-term reliability requirement. Motivated by this observation,
we develop a reliability-constrained predictive monitoring
framework that combines probabilistic grant modeling with
modem-level power-state control. The main contributions of
this paper can be summarized as follows:
\begin{itemize}
    \item We establish a reliability-constrained IoT-D-side PDCCH monitoring optimization framework for 5G NR terminals. Different from pure grant prediction, the proposed formulation directly optimizes the monitoring decision by jointly considering receiver energy consumption and missed-grant reliability.
    \item We construct a modem-level energy objective that includes receiver operating energy, component switching energy, and prediction-computation energy. This is consistent with the uploaded paper’s power modeling philosophy, where receiver-chain energy and FLOP-based algorithm overhead are both considered in the evaluation framework.
    \item We introduce an FNR-based reliability constraint to prevent overly aggressive sleeping decisions. The proposed formulation therefore avoids two trivial solutions: always monitoring the PDCCH, which wastes energy, and always skipping monitoring, which causes unacceptable missed grants.
   \item We propose an MoE-IOHMM-based predictive monitoring scheme to handle heterogeneous grant-arrival patterns across different applications and traffic conditions. Specifically, multiple IO-HMM experts are constructed to capture distinct hidden scheduling dynamics, while a gating network adaptively routes the current IoT-D-side historical observation to the most suitable experts. The aggregated grant probability is further converted into a PDCCH monitoring action through a reliability-aware threshold controller, thereby linking expert-based grant prediction with FNR-constrained modem power control.
\end{itemize}

\section{Related Work}
Energy-efficient operation of IoT-D has become an important research topic in 5G NR systems, since the cellular modem and receiver chain consume non-negligible energy during continuous control-channel monitoring. In particular, the IoT-D has to blindly monitor the PDCCH to obtain downlink control information, even when no valid scheduling grant is present. This motivates predictive DPM, where the receiver chain can be switched into a low-power state when the upcoming TTI is predicted to be grant-free. The uploaded thesis also identifies continuous PDCCH blind detection as a major source of unnecessary IoT-D energy consumption and points out that predictive DPM can reduce receiver activity, but only under controlled missed-grant risk.

\subsection{DRX-Based and Network-Side Power Saving}
A large body of work has studied IoT-D energy saving from the perspective of discontinuous reception (DRX), wake-up signaling, and network-side configuration. \cite{b1} proposed a comprehensive survey of DRX mechanisms from LTE and 5G NR to 6G, summarizing how sleep/active cycles reduce IoT-D power consumption under latency and reliability constraints. \cite{b2} analyzed 3GPP 5G NR power-saving techniques with a focus on their latency and reliability implications. \cite{b3} developed an accurate analytical model for LTE DRX with predetermined DRX cycles, while \cite{b4} compared adjustable and fixed DRX mechanisms for LTE/LTE-Advanced power saving. Wake-up-radio-based mobile access was further investigated in \cite{b5}, where the IoT-D can remain in a low-power state until a wake-up indication is received. Another line of work improves power saving through network-side traffic prediction or resource control. \cite{b6} proposed an enhanced DRX mechanism for 5G power saving. \cite{b7} studied energy and resource efficiency using traffic prediction and classification in cellular networks, and \cite{b8} proposed a base-station sleeping strategy in heterogeneous cellular networks based on user traffic prediction. \cite{b9} investigated traffic-driven sounding reference signal resource allocation in beyond-5G networks. These methods mainly optimize base-station-side configuration, DRX parameters, wake-up signaling, or network resource allocation. However, they usually require network-side coordination or additional signaling, whereas the problem considered in this paper focuses on IoT-D-side PDCCH monitoring decisions based only on locally observable historical information.

\subsection{IoT-D-Side Predictive Dynamic Power Management}
Different from network-side DRX optimization, IoT-D-side predictive DPM attempts to predict whether a grant will appear and then determines whether the receiver chain should remain active. \cite{b10} developed an end-to-end power estimation framework for heterogeneous cellular LTE SoCs, providing an early basis for modem-level power modeling. \cite{b11} proposed adaptive predictive power management for mobile LTE devices, and \cite{b12} studied grant-prediction-based DPM for 5G mobile devices. \cite{b13} modeled LTE scheduling as a binary time series for machine-learning-based prediction. \cite{b14} and \cite{b15} further proposed clustering-based scenario-aware LTE grant prediction and multi-step-ahead grant prediction for cellular modem DPM. These studies show that scheduling grants exhibit temporal structure and can be exploited for receiver power control. Machine-learning-based grant prediction has also been considered. \cite{b16} proposed a 1DCNN-TCN-GRU hybrid model for network traffic grant classification, combining convolutional and recurrent temporal feature extraction. The uploaded thesis also summarizes prior IoT-D-side methods based on multilayer perceptrons, random forests, k-nearest neighbors, clustering, support vector machines, and random forests for grant prediction, and notes that multi-step prediction can reduce receiver energy by exploiting historical scheduling information. However, these methods often focus on prediction accuracy rather than explicitly formulating reliability-constrained energy minimization. In sparse-grant scenarios, accuracy alone can be misleading: a model may achieve high accuracy by predicting most TTIs as grant-free, while still causing unacceptable false negatives. This motivates our FNR-constrained formulation.

\subsection{Reinforcement Learning for Power Management}
Reinforcement learning has been widely used for sequential power-management problems. \cite{b17} derived near-optimal power management policies using model-free reinforcement learning and Bayesian classification. General reinforcement learning techniques for energy systems were surveyed in \cite{b18}. In the IoT-D-side DPM context, the uploaded thesis designs an Actor-Critic method that treats monitoring control as a partially observable sequential decision problem, where the state is formed by stacked historical observations and the policy outputs monitoring-window actions. The reward function assigns a larger negative reward to missed grants so that the learned policy prioritizes reliability while still exploiting sleeping opportunities. Although reinforcement learning provides an end-to-end decision-making tool, it may require substantial training data and can suffer from unstable exploration, especially when false-negative events are rare but costly. In contrast, probabilistic sequence models can provide interpretable grant probabilities, which are easier to combine with explicit reliability-aware threshold control. Therefore, instead of directly relying on an end-to-end DRL policy, our design preserves the probabilistic modeling advantage of IO-HMM and further improves its adaptability using MoE routing.

\subsection{IO-HMM and Sequential Grant Modeling}
Hidden Markov models and their extensions are widely used to model sequential processes with latent states. For grant prediction, the hidden state can represent the unobservable scheduling phase of the gNB-side MAC scheduler, while the observed output corresponds to whether a valid grant appears. \cite{b19,b20,b21} introduced input-output hidden Markov models for sequence processing, where transition and emission distributions can depend on external inputs. This is suitable for PDCCH grant prediction because protocol-side events such as scheduling requests and CDRX state transitions can affect subsequent grant arrivals. The uploaded thesis uses a high-order IO-HMM to model the grant-generation process under partial observability. Specifically, it observes that real 5G NR grant sequences are affected by application behavior, protocol mechanisms, and network environments, and are therefore not purely random. It further transforms a high-order IO-HMM into an equivalent first-order composite state representation and jointly optimizes the model order and monitoring-window parameters. This provides a strong probabilistic basis for IoT-D-side predictive monitoring. Nevertheless, a single IO-HMM may still be insufficient when grant patterns vary significantly across applications, scenarios, and traffic modes. This motivates introducing multiple IO-HMM experts rather than forcing all traffic patterns into one global model.

\section{System Model}

\begin{figure*}[htbp]
  \centering
  \includegraphics[width=0.75\textwidth, height=0.35\textwidth]{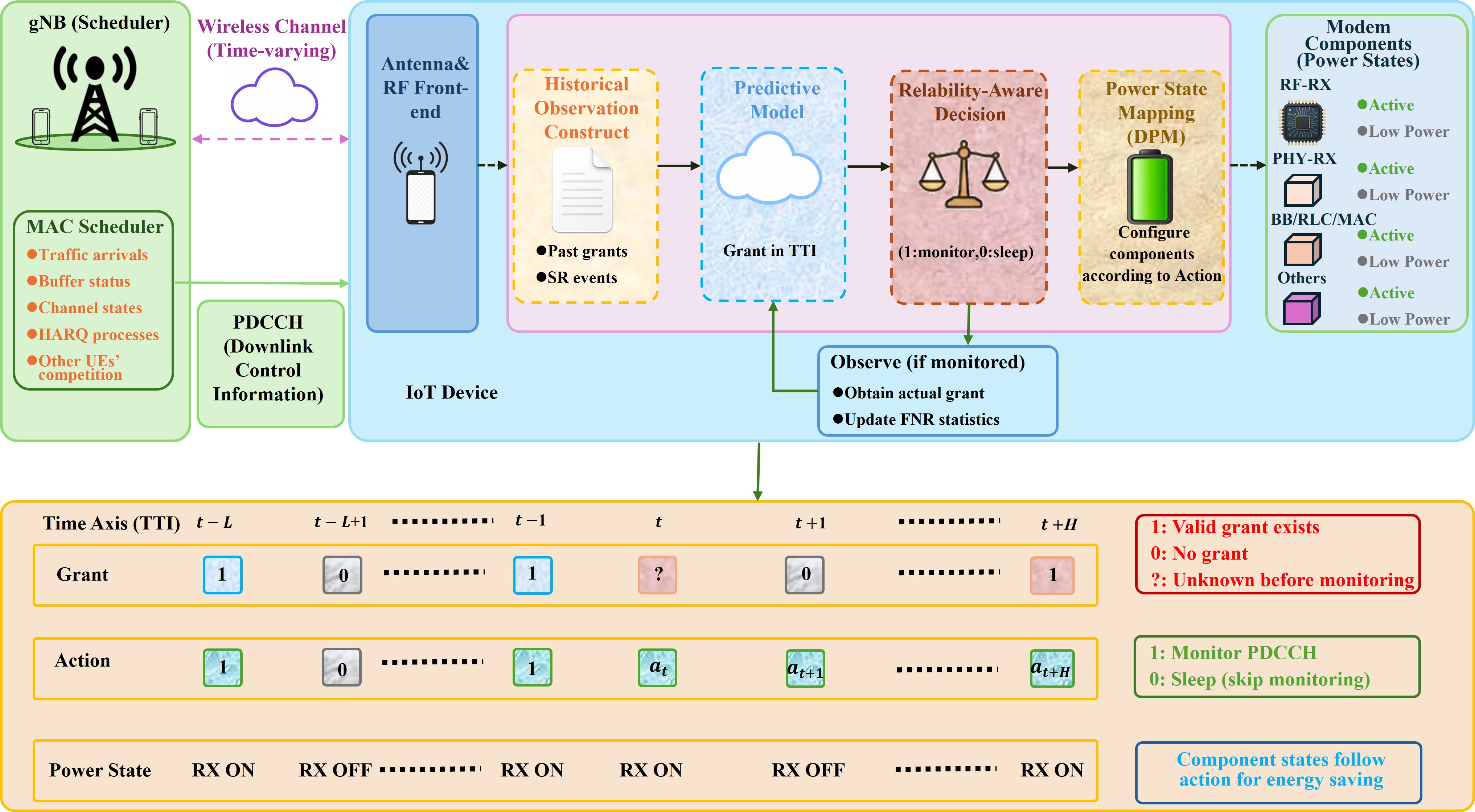}
  \captionsetup{justification=centering}
  \caption{System model of reliability-constrained IoT-D-side predictive power scheduling for PDCCH monitoring. The IoT-D exploits historical scheduling observations, such as past grants, MCS, TBS, inter-grant interval, and grant density, to decide whether to monitor the PDCCH in the current slot. The decision jointly affects modem component power states, slot-level energy cost, and missed-grant events, thereby creating an energy-reliability tradeoff under the FNR constraint.}\vspace{-10pt}
\label{FIGURE1_0}
\end{figure*}

\subsection{5G NR Downlink Scheduling and PDCCH Monitoring Model}

We consider a 5G NR downlink system consisting of one gNB and one IoT-D. The system operates in a slotted manner, where each slot, or TTI, is indexed by $t \in \{1,2,\ldots,T\}$. At the beginning of each TTI, the gNB-side MAC scheduler determines whether to allocate downlink or uplink transmission resources to the IoT-D. The scheduling decision is conveyed through the PDCCH, where the DCI specifies the corresponding scheduling grant, including the resource allocation, modulation and coding scheme, and HARQ-related information. Since the IoT-D does not know in advance whether a valid scheduling grant will appear in the current TTI, it must perform blind PDCCH detection over its configured search space. If a valid DCI addressed to the IoT-D is decoded, the IoT-D proceeds to receive the corresponding physical downlink shared channel (PDSCH) or performs the instructed uplink transmission. Otherwise, if no valid scheduling grant is detected, the receiver chain can be switched to a low-power state during the remaining part of the TTI. This blind monitoring mechanism is necessary for reliable control-channel reception, but it also leads to unnecessary energy consumption when no grant is present. The uploaded paper explicitly treats this continuous PDCCH monitoring process as a key source of IoT-D-side energy waste and motivates predictive DPM to reduce such redundant receiver activity. Let $G_t \in \{0,1\}$ denote the actual scheduling-grant state in TTI $t$, where $G_t=1$ indicates that a valid scheduling grant for the IoT-D exists, and $G_t=0$ indicates that no valid grant exists. The sequence $\{G_t\}_{t=1}^{T}$ represents the grant-arrival process observed from the IoT-D side. Importantly, $G_t$ is not known before the IoT-D makes its monitoring decision. The IoT-D can only infer the likelihood of grant arrival from historical observations and local protocol events. The grant-arrival process is not assumed to be independent across TTIs. In practical 5G NR systems, the scheduling decision depends on the gNB-side MAC scheduler, which jointly considers traffic demand, channel quality, buffer status, QoS requirements, retransmission status, and multi-user competition. These factors are not fully observable to the IoT-D. Therefore, the grant sequence is driven by hidden scheduling dynamics and exhibits temporal correlation.

\subsection{IoT-D-Side Historical Observation Model}
At the IoT-D side, the internal state of the gNB scheduler is unavailable. The IoT-D can only use locally observable information obtained from previous TTIs. Let $\mathcal{H}_t$ denote the IoT-D-side historical observation available before making the monitoring decision in TTI $t$. In general, $\mathcal{H}_t$ may include the previously observed grant indicators, PDCCH decoding results, scheduling-request-related events, local CDRX state transitions, and other modem-side features that are available before the current scheduling outcome is revealed. We introduce an input feature vector
\begin{align}
    \mathbf{u}_t =
    \left[
    SR_t,\;
    \exp(-\alpha \Delta SR_t),\;
    C_t^{\mathrm{rise}},\;
    \exp(-\alpha \Delta C_t^{\mathrm{rise}})
    \right]^T,\label{p1}
\end{align}
where $SR_t$ indicates whether a scheduling request occurs at time $t$, $\Delta SR_t$ denotes the elapsed time since the most recent SR event, $C_t^{\mathrm{rise}}$ indicates whether the local CDRX state machine has a rising edge from sleep to active state at time $t$, and $\Delta C_t^{\mathrm{rise}}$ denotes the elapsed time since the most recent CDRX rising edge. The constant $\alpha>0$ controls the exponential decay of event influence over time. This feature design reflects two important properties of the scheduling process. First, an SR event indicates that new uplink traffic has arrived at the IoT-D side and may trigger subsequent uplink grants or related downlink control interactions. The uploaded paper reports that the conditional probability of observing a valid downlink grant increases significantly after an SR event and then gradually decays over time, which motivates the use of both $SR_t$ and $\Delta SR_t$ as predictive features. Second, the CDRX mechanism imposes a protocol-level structure on IoT-D monitoring behavior. Since the IoT-D periodically transitions from sleep to active state to monitor possible scheduling grants, the local CDRX rising edge provides useful timing information for grant prediction.  For compact notation, the history used for decision making can be written as
\begin{align}
\mathcal{H}_t =
    \left\{
    G_{t-L}, \ldots, G_{t-1},
    \mathbf{u}_{t-L}, \ldots, \mathbf{u}_{t}
    \right\},\label{p2}    
\end{align}
where $L$ is the historical observation length. When the IoT-D skips PDCCH monitoring in a certain TTI, the true grant state in that TTI may not be directly observed. In this case, the historical sequence may contain incomplete or decision-dependent observations, which further reinforces the partial observability of the system.

\subsection{Hidden Scheduling-State Model}
To characterize the fact that grant arrivals are driven by unobservable gNB-side scheduling dynamics, we introduce a latent scheduling state $S_t \in \mathcal{S}$, where $\mathcal{S}$ denotes a finite hidden-state space. The hidden state $S_t$ abstracts the unobservable scheduling phase of the IoT-D, such as traffic burst status, scheduler preference, HARQ-related activity, or implicit QoS-driven priority. The IoT-D does not observe $S_t$ directly; it only observes the grant outcome when PDCCH is monitored and the local input features $\mathbf{u}_t$. The hidden state evolves over time according to a controlled or input-dependent Markovian transition structure. A general form can be written as
\begin{align}
\Pr(S_t=s' \mid S_{t-1}=s,\mathbf{u}_t)=A_{s,s'}(\mathbf{u}_t),\label{p3}
\end{align}
where $A_{s,s'}(\mathbf{u}_t)$ denotes the transition probability from hidden state $s$ to hidden state $s'$ conditioned on the current input feature vector. The actual grant state is then generated according to an emission distribution
\begin{align}
\Pr(G_t=g \mid S_t=s,\mathbf{u}_t)=B_s(g \mid \mathbf{u}_t),\label{p4}
\end{align}
where $g\in\{0,1\}$. This input-dependent hidden-state model is consistent with the IO-HMM interpretation in the uploaded paper. 
For a higher-order model, the hidden state transition may depend on multiple previous hidden states and input events. A $K$-order representation can be written as
\begin{align}
    \Pr(S_t \mid S_{t-1},\ldots,S_{t-K},
    \mathbf{u}_t,\ldots,\mathbf{u}_{t-K+1}),\label{p5}
\end{align}
where $K$ denotes the memory order. Equivalently, the high-order hidden process can be transformed into a first-order process by defining a composite state
\begin{align}
    \widetilde{S}_t =
    (S_t,S_{t-1},\ldots,S_{t-K+1}).\label{p6}
\end{align}
This representation preserves the long-term dependency while allowing the process to be handled using a first-order state-space form. In the system model, this hidden-state abstraction is used only to describe the underlying grant-generation mechanism; the final monitoring decision is still made at the IoT-D based on observable history.

\subsection{PDCCH Monitoring Action and Receiver Operation}

At each TTI $t$, before observing $G_t$, the IoT-D chooses a PDCCH monitoring action $a_t \in \{0,1\}$, where $a_t=1$ means that the IoT-D monitors the PDCCH in TTI $t$, and $a_t=0$ means that the IoT-D skips PDCCH monitoring and switches the receiver chain into a low-power state. If $a_t=1$, the IoT-D activates the necessary receiver components, performs blind PDCCH decoding, and observes whether a valid grant exists. In this case, if $G_t=1$, the IoT-D successfully obtains the scheduling information and proceeds with the corresponding data reception or uplink transmission procedure. If $G_t=0$, the IoT-D discovers that no valid grant exists after PDCCH monitoring, and the receiver components may be turned off during the remaining part of the TTI. If $a_t=0$, the IoT-D does not monitor the PDCCH in TTI $t$. This reduces receiver activity and saves energy. However, if a valid grant actually exists in that TTI, i.e., $G_t=1$, the grant cannot be decoded by the IoT-D. This event is referred to as a missed grant or false negative event. We define the missed-grant indicator as $M_t = G_t(1-a_t)$. Thus, $M_t=1$ only when a valid grant exists but the IoT-D skips PDCCH monitoring. The action $a_t$ can be generated by a monitoring policy $a_t = \pi(\mathcal{H}_t)$, where $\pi(\cdot)$ maps the IoT-D-side historical observation to a binary monitoring action. In a prediction-based implementation, the IoT-D may first estimate the grant probability
$p_t = \Pr(G_t=1 \mid \mathcal{H}_t)$, and then determine $a_t$ according to the predicted grant likelihood. The specific design of $\pi(\cdot)$ is left to the solution methodology section, while this subsection only defines the monitoring action and its operational consequence.

\subsection{Receiver Component Power-State Model}

The IoT-D receiver chain consists of multiple modem components, such as the radio-frequency receiver, physical-layer receiver, and baseband processing modules. Let $\mathcal{M}$ denote the set of receiver-related components, and let $q_{m,t} \in \{0,1\}$ denote the power state of component $m\in\mathcal{M}$ in TTI $t$. Specifically,
$q_{m,t}=1$ means that component $m$ is active, while $q_{m,t}=0$
means that component $m$ is in a low-power or off state.
For PDCCH monitoring, a subset of components must be active. Let $\mathcal{M}_{\mathrm{PDCCH}} \subseteq \mathcal{M}$ denote the set of components required for PDCCH blind detection, such as RF-RX and PHY-RX. The monitoring action and component states must satisfy the operational relation
\begin{align}
    a_t \leq q_{m,t}, \forall m\in\mathcal{M}_{\mathrm{PDCCH}}.\label{p7}
\end{align}
This means that if the IoT-D decides to monitor the PDCCH, all required receiver components must be active. 

\subsection{Energy Consumption Model}

The energy consumption in each TTI is composed of receiver operating energy, component switching energy, and prediction-computation energy. Let $E_t$
denote the total IoT-D-side energy consumption associated with PDCCH monitoring and power-state control in TTI $t$. We decompose it as
\begin{align}
    E_t = E_t^{\mathrm{op}} + E_t^{\mathrm{sw}} + E_t^{\mathrm{cmp}}.\label{p8}
\end{align}
The receiver operating energy is determined by the active or low-power state of each component. It can be expressed as
\begin{align}
    E_t^{\mathrm{op}}
    =
    \Delta_t
    \sum_{m\in\mathcal{M}}
    \left[
    P_m^{\mathrm{on}} q_{m,t}
    +
    P_m^{\mathrm{off}}(1-q_{m,t})
    \right],\label{p9}
\end{align}
where $\Delta_t$ is the duration of one TTI, $P_m^{\mathrm{on}}$ is the active-state power of component $m$, and $P_m^{\mathrm{off}}$ is the low-power-state power of component $m$. The component switching energy captures the energy overhead caused by power-state transitions between adjacent TTIs. It is modeled as
\begin{align}
    E_t^{\mathrm{sw}}
    =
    \sum_{m\in\mathcal{M}}
    C_m^{\mathrm{sw}}
    \left|q_{m,t}-q_{m,t-1}\right|,\label{p10}
\end{align}
where $C_m^{\mathrm{sw}}$ denotes the energy cost of switching component $m$ between active and low-power states. The prediction-computation energy accounts for the energy consumed by running the grant prediction or monitoring decision algorithm. Following the FLOP-based evaluation principle used in \cite{b010}, this term can be written as $E_t^{\mathrm{cmp}}=\kappa \mathrm{FLOPs}_t$,
where $\mathrm{FLOPs}_t$ is the number of floating-point operations required by the prediction algorithm in TTI $t$, and $\kappa$ is the energy cost per floating-point operation.

\subsection{False Negative Event and Reliability Indicator}
Because predictive DPM may skip PDCCH monitoring, reliability must be characterized by the occurrence of missed grants. As defined above, the missed-grant indicator is
\begin{align}
    M_t = G_t(1-a_t).\label{p11}
\end{align}
If $G_t=1$ and $a_t=0$, then $M_t=1$, meaning that a valid scheduling grant is missed. If $a_t=1$, then $M_t=0$, because the IoT-D monitors the PDCCH and can detect the grant if it is present. Over a finite horizon $T$, the empirical false negative rate is
\begin{align}
    \mathrm{FNR}_T
    =
    \frac{\sum_{t=1}^{T} M_t}
    {\sum_{t=1}^{T} G_t+\delta}
    =
    \frac{\sum_{t=1}^{T} G_t(1-a_t)}
    {\sum_{t=1}^{T} G_t+\delta},\label{p12}
\end{align}
where $\delta>0$ is a small constant used to avoid division by zero when no grant appears in the observation horizon. This quantity measures the fraction of valid grants that are missed due to skipped PDCCH monitoring. The FNR is an essential reliability indicator for IoT-D-side predictive DPM. A lower FNR indicates that the IoT-D rarely misses valid grants, while a higher FNR implies more aggressive sleeping and potentially degraded communication reliability. 

The system model has three important characteristics. First, the scheduling process is partially observable. The IoT-D does not know the gNB-side scheduling state, the traffic queues of other users, the scheduler decision metric, or the exact resource allocation logic. It only observes local historical information and decoded control-channel outcomes. Second, the grant-arrival process is temporally correlated. Real grant traces are affected by traffic bursts, SR events, CDRX cycles, and scheduler behavior. Therefore, a memoryless decision rule is insufficient for reliable monitoring control. This motivates the use of historical observation sequences and hidden-state models. Third, the monitoring decision directly affects both energy consumption and observation quality. Monitoring PDCCH increases energy consumption but provides reliable grant observation. Skipping PDCCH saves receiver energy but may create missed grants and incomplete historical observations. Therefore, IoT-D-side PDCCH monitoring is naturally a sequential decision process with coupled energy and reliability effects. This system model provides the basis for the subsequent problem formulation, where the monitoring action and receiver component states will be optimized under a reliability requirement.

\section{Problem Formulation}
Based on the system model described above, we formulate the
IoT-D-side PDCCH monitoring design as a reliability-constrained
long-term energy minimization problem. At each TTI, the IoT-D
decides whether to monitor the PDCCH before the actual
scheduling-grant state is observed. Monitoring the PDCCH
guarantees reliable grant detection but keeps the receiver chain
active and therefore consumes modem energy. In contrast,
skipping PDCCH monitoring allows the IoT-D to switch receiver
components into low-power states, but may cause a missed
grant when a valid scheduling grant is actually present. Hence,
the key design objective is to minimize the long-term IoT-D-side
energy consumption while ensuring that the missed-grant
probability, characterized by the FNR,
remains below a prescribed reliability threshold. This
energy-reliability tradeoff is consistent with the predictive
PDCCH monitoring framework, where receiver operating
energy, switching energy, prediction-computation energy, and
missed-grant events are jointly considered.

Accordingly, the reliability-constrained PDCCH monitoring
problem is formulated as
\begin{subequations}
\begin{align}
\min_{\{a_t,q_{m,t}\}}
&
\limsup_{T\rightarrow\infty}
\frac{1}{T}
\mathbb{E}
\left[
\sum_{t=1}^{T}
\left(
E_t^{\mathrm{op}}
+
E_t^{\mathrm{sw}}
+
E_t^{\mathrm{cmp}}\right.\right.\nonumber\\
&\left.\left.+
\lambda G_t(1-a_t)
\right)
\right]
\label{p13a}
\\
\mathrm{s.t.}
&
\limsup_{T\rightarrow\infty}
\mathbb{E}
\left[
\frac{\sum_{t=1}^{T}G_t(1-a_t)}
{\sum_{t=1}^{T}G_t+\delta}
\right]
\leq \epsilon,
\label{p13b}
\\
&
a_t \leq q_{m,t},
\quad \forall m\in\mathcal{M}_{\mathrm{PDCCH}},\ \forall t,
\label{p13c}
\\
&
a_t\in\{0,1\},\quad \forall t,
\label{p13d}
\\
&
q_{m,t}\in\{0,1\},
\quad \forall m\in\mathcal{M},\ \forall t.
\label{p13e}
\end{align}\label{p13}%
\end{subequations}
In \eqref{p13a}, the objective function minimizes the
long-term average expected IoT-D-side cost. The terms
$E_t^{\mathrm{op}}$, $E_t^{\mathrm{sw}}$, and
$E_t^{\mathrm{cmp}}$ denote the receiver operating energy,
component switching energy, and prediction-computation
energy, respectively. The additional term
$\lambda G_t(1-a_t)$ penalizes missed grants, where
$\lambda>0$ controls the reliability cost of skipping PDCCH
monitoring when a valid grant exists. Constraint \eqref{p13b} imposes the long-term reliability
requirement. Since $G_t(1-a_t)=1$ only when the IoT-D skips
PDCCH monitoring while a valid scheduling grant is present,
the numerator represents the total number of missed grants.
The denominator represents the total number of actual valid
grants, with a small constant $\delta>0$ added to avoid
division by zero. Therefore, this constraint ensures that the
long-term FNR does not exceed the prescribed threshold
$\epsilon$. Constraint \eqref{p13c} describes the operational
coupling between the monitoring decision and the receiver
component states. If the IoT-D decides to monitor the PDCCH,
i.e., $a_t=1$, then every component required for PDCCH blind
detection, such as RF-RX and PHY-RX, must be active. This
is enforced by $a_t\leq q_{m,t}$ for all
$m\in\mathcal{M}_{\mathrm{PDCCH}}$. Constraints \eqref{p13d} and \eqref{p13e} specify the
binary nature of the decision variables. The variable $a_t$
indicates whether the IoT-D monitors or skips the PDCCH in TTI
$t$, while $q_{m,t}$ represents whether modem component
$m$ is active or in a low-power state. Problem $\mathbf{P1}$ is a constrained partially observable
sequential decision problem. The difficulty lies in the fact that
the IoT-D must choose $a_t$ before observing $G_t$, while the
grant-arrival process is driven by the hidden gNB-side
scheduler and exhibits temporal correlation. Therefore, an
effective solution requires a predictive model that can infer
the grant-arrival probability from historical IoT-D-side
observations and then convert this probability into a
reliability-aware monitoring action.

\section{Proposed MoE-IOHMM Reliability-Aware Predictive Monitoring Scheme}
To solve the reliability-constrained IoT-D-side PDCCH monitoring problem, we propose a mixture-of-experts input-output hidden Markov model, termed the \textbf{MoE-IOHMM predictive monitoring scheme}. The proposed method is designed to preserve the probabilistic interpretability of IO-HMM while improving its adaptability to heterogeneous scheduling-grant patterns across different applications and radio environments. The main idea is to construct several IO-HMM experts, each of which captures one possible hidden scheduling mode behind the observed grant sequence. A gating network dynamically assigns weights to different experts according to the current IoT-D-side historical observation. The weighted expert outputs are then combined into a final grant probability, which is further converted into a PDCCH monitoring action through an FNR-aware threshold controller.

\subsection{Historical Observation and Input Construction}
At TTI $t$, the IoT-D constructs the historical observation as
\begin{align}
\mathcal{H}_t =
\left\{
G_{t-L},\ldots,G_{t-1},
\mathbf{U}_{t-L},\ldots,\mathbf{U}_{t}
\right\},\label{p14}%
\end{align}
where $L$ is the observation window length, $G_t \in \{0,1\}$ is the effective grant indicator, and $\mathbf{U}_t$ denotes the local input feature vector available before the current grant state is observed.
Following the uploaded paper, the input feature vector is defined as
\begin{align}
\mathbf{U}_t =
\left[
SR_t,\,
\exp(-\alpha \Delta SR_t),\,
C_t^{\mathrm{rise}},\,
\exp(-\alpha \Delta C_t^{\mathrm{rise}})
\right]^T .\label{p15}%
\end{align}
Here, $SR_t$ indicates whether a scheduling request occurs at TTI $t$, $\Delta SR_t$ is the elapsed time since the most recent SR event, $C_t^{\mathrm{rise}}$ denotes the CDRX rising edge from sleep to active state, and $\Delta C_t^{\mathrm{rise}}$ is the elapsed time since the latest CDRX rising edge. The exponential decay terms encode the gradually weakening influence of SR/CDRX events over time. This feature design is directly consistent with the paper, where SR and CDRX rising-edge features are introduced to capture protocol-driven grant correlations and reduce missed detections. For a high-order IO-HMM expert, we further define the stacked input as
\begin{align}
\mathbf{V}_t =
\left[
\mathbf{U}_t^T,\,
\mathbf{U}_{t-1}^T,\,
\ldots,\,
\mathbf{U}_{t-K+1}^T
\right]^T ,\label{p16}%
\end{align}
where $K$ is the memory order. This allows the model to capture the lasting influence of recent protocol events.

\subsection{IO-HMM Expert Model}

The proposed MoE-IOHMM contains $N_E$ experts. The $e$-th expert is an IO-HMM parameterized by
\begin{align}
\Theta_e =
\left\{
\boldsymbol{\pi}^{(e)},
\mathbf{A}^{(e)},
\mathbf{B}^{(e)}
\right\}, e=1,\ldots,N_E,\label{p17}%
\end{align}
where $\boldsymbol{\pi}^{(e)}$ is the initial hidden-state distribution, $\mathbf{A}^{(e)}$ is the input-dependent transition model, and $\mathbf{B}^{(e)}$ is the emission model. Let the basic hidden scheduling state of expert $e$ be $\widetilde{S}_t^{(e)} \in \{0,1\}$, where $\widetilde{S}_t^{(e)}=0$ represents an inactive or low-grant scheduling phase, while $\widetilde{S}_t^{(e)}=1$ represents an active or grant-intensive scheduling phase. This is consistent with the uploaded paper, where the hidden state is used to abstract IoT-D-invisible gNB-side scheduling activity rather than a directly observable protocol field. To capture long-term dependency, the high-order hidden state is rewritten as a first-order composite state:
\begin{align}
S_t^{(e)}
=
\left(
\widetilde{S}_t^{(e)},
\widetilde{S}_{t-1}^{(e)},
\ldots,
\widetilde{S}_{t-K+1}^{(e)}
\right).\label{p18}%
\end{align}
Thus, although the original process has $K$-order memory, the composite state $S_t^{(e)}$ follows a first-order Markov structure.
For expert $e$, the transition probability is modeled as
\begin{align}
A_{ij}^{(e)}(\mathbf{V}_t)
=
\Pr\left(
S_t^{(e)}=j
\mid
S_{t-1}^{(e)}=i,\mathbf{V}_t
\right).\label{p19}%
\end{align}
A practical parameterization is the softmax transition model:
\begin{align}
A_{ij}^{(e)}(\mathbf{V}_t)
=
\frac{
\exp\left(
\alpha_{ij}^{(e)}
+
\mathbf{a}_{ij}^{(e)T}\mathbf{V}_t
\right)
}{
\sum_{j'}
\exp\left(
\alpha_{ij'}^{(e)}
+
\mathbf{a}_{ij'}^{(e)T}\mathbf{V}_t
\right)
}.\label{p20}%
\end{align}
This means the transition between hidden scheduling phases is not fixed, but depends on SR/CDRX-driven input features. The emission probability is defined as
\begin{align}
B_j^{(e)}(g|\mathbf{V}_t)
=
\Pr\left(
G_t=g
\mid
S_t^{(e)}=j,\mathbf{V}_t
\right),
\quad g\in\{0,1\}.\label{p21}%
\end{align}
For binary grant output, the grant-present probability can be parameterized by a logistic model:
\begin{align}
&B_j^{(e)}(1|\mathbf{V}_t)
=
\sigma\left(
b_j^{(e)}
+
\mathbf{b}_j^{(e)T}\mathbf{V}_t
\right),\nonumber\\
&B_j^{(e)}(0|\mathbf{V}_t)
=
1-
B_j^{(e)}(1|\mathbf{V}_t),\label{p22}%
\end{align}
where $\sigma(x)=\frac{1}{1+\exp(-x)}$. Therefore, each IO-HMM expert gives a probabilistic description of how hidden scheduling states evolve and how grants are emitted under local IoT-D-side inputs.

\subsection{Forward Prediction of Each IO-HMM Expert}
For expert $e$, define the filtered belief over hidden composite states after observing historical information up to $t-1$ as
\begin{align}
\alpha_{t-1}^{(e)}(i)
=
\Pr\left(
S_{t-1}^{(e)}=i
\mid
\mathcal{H}_{t-1};\Theta_e
\right).\label{p23}%
\end{align}
Before observing $G_t$, the one-step predicted hidden-state belief is
\begin{align}
\widetilde{\alpha}_{t}^{(e)}(j)
=
\sum_i
\alpha_{t-1}^{(e)}(i)
A_{ij}^{(e)}(\mathbf{V}_t).\label{p24}%
\end{align}
Then, the $e$-th expert predicts the grant probability as
\begin{align}
p_t^{(e)}
=
\Pr\left(
G_t=1
\mid
\mathcal{H}_t;\Theta_e
\right)
=
\sum_j
\widetilde{\alpha}_{t}^{(e)}(j)
B_j^{(e)}(1|\mathbf{V}_t).\label{p25}%
\end{align}
This is the key prediction equation. It means that each expert first predicts the current hidden scheduling state and then marginalizes over hidden states to obtain the probability of a valid scheduling grant. After $G_t$ is observed during training or evaluation, the filtering belief can be updated as
\begin{align}
\alpha_t^{(e)}(j)
=
\frac{
B_j^{(e)}(G_t|\mathbf{V}_t)
\sum_i
\alpha_{t-1}^{(e)}(i)
A_{ij}^{(e)}(\mathbf{V}_t)
}{
\sum_{j'}
B_{j'}^{(e)}(G_t|\mathbf{V}_t)
\sum_i
\alpha_{t-1}^{(e)}(i)
A_{ij'}^{(e)}(\mathbf{V}_t)
}.\label{p26}%
\end{align}
This recursion is the probabilistic core of the IO-HMM expert.

\subsection{Gating Network and MoE Aggregation}

A single IO-HMM may be insufficient because the grant sequence can vary significantly across different applications and scenarios. Therefore, we introduce a gating network to combine multiple IO-HMM experts. Given $\mathcal{H}_t$, the gating network outputs
\begin{align}
\boldsymbol{\omega}_t
=
\left[
\omega_t^{(1)},\omega_t^{(2)},\ldots,\omega_t^{(N_E)}
\right],\label{p27}%
\end{align}
, and
\begin{align}
\omega_t^{(e)}
=
\frac{
\exp\left(r_e(\mathcal{H}_t;\boldsymbol{\phi})\right)
}{
\sum_{\ell=1}^{N_E}
\exp\left(r_\ell(\mathcal{H}_t;\boldsymbol{\phi})\right)
},\label{p28}%
\end{align}
where $\omega_t^{(e)} \geq 0$,$\sum_{e=1}^{N_E} \omega_t^{(e)} = 1$. Here, $\boldsymbol{\phi}$ denotes the gating network parameters, and $r_e(\mathcal{H}_t;\boldsymbol{\phi})$ is the score assigned to expert $e$. The final MoE-IOHMM grant probability is $p_t
=\sum_{e=1}^{N_E}
\omega_t^{(e)} p_t^{(e)}$. This equation is the MoE aggregation rule. The gating network adaptively selects the experts according to the current historical pattern. For instance, when SR has occurred recently and the grant probability is expected to increase, experts that model SR-triggered grant dynamics can receive larger weights. When the sequence shows long inactive intervals, sparse-traffic experts may dominate.

\subsection{Training Objective of MoE-IOHMM}
The MoE-IOHMM predictor is trained using historical traces. Since valid grants can be sparse, directly minimizing standard prediction error may bias the model toward predicting no grant. To reduce missed grants, we adopt a reliability-weighted negative log-likelihood. For each sample at TTI $t$, the likelihood of the observed grant is
\begin{align}
\Pr(G_t|\mathcal{H}_t)
=
p_t^{G_t}
(1-p_t)^{1-G_t}.,\label{p29}%
\end{align}
The weighted prediction loss is
\begin{align}
\mathcal{L}_{\mathrm{pred}}
=
-
\sum_{t=1}^{T}
\left[
\chi G_t \log p_t
+
(1-G_t)\log(1-p_t)
\right],,\label{p30}%
\end{align}
where $\chi>1$ is the reliability weight for grant-present samples. A larger $\chi$ gives more importance to $G_t=1$ samples and reduces the risk of false negative predictions. To encourage specialization among experts, we introduce the posterior responsibility of expert $e$:
\begin{align}
\gamma_t^{(e)}
=
\frac{
\omega_t^{(e)}
\Pr(G_t|\mathcal{H}_t;\Theta_e)
}{
\sum_{\ell=1}^{N_E}
\omega_t^{(\ell)}
\Pr(G_t|\mathcal{H}_t;\Theta_\ell)
}.\label{p31}%
\end{align}
Here,
\begin{align}
\Pr(G_t|\mathcal{H}_t;\Theta_e)
=
\left(p_t^{(e)}\right)^{G_t}
\left(1-p_t^{(e)}\right)^{1-G_t}.\label{p32}%
\end{align}
The responsibility $\gamma_t^{(e)}$ measures how much expert $e$ explains the observed grant outcome at time $t$. Experts with larger responsibility receive stronger parameter updates. The expert-specific weighted loss can be written as
\begin{align}
\mathcal{L}_e
=
-
\sum_{t=1}^{T}
\gamma_t^{(e)}
\left[
\chi G_t \log p_t^{(e)}
+
(1-G_t)\log(1-p_t^{(e)})
\right].\label{p33}%
\end{align}
The overall training objective is
\begin{align}
\min_{\{\Theta_e\},\boldsymbol{\phi}}
\sum_{e=1}^{N_E}
\mathcal{L}_e
+
\lambda_g \mathcal{L}_{\mathrm{gate}},\label{p34}%
\end{align}
where $\mathcal{L}_{\mathrm{gate}}$ is used to train the gating network. A simple gating loss is
\begin{align}
\mathcal{L}_{\mathrm{gate}}
=
-
\sum_{t=1}^{T}
\sum_{e=1}^{N_E}
\gamma_t^{(e)}
\log \omega_t^{(e)}.\label{p35}%
\end{align}
This loss encourages the gating network to assign higher weights to experts that better explain the current observation pattern. To avoid expert collapse, where only one expert dominates all samples, an entropy regularization term can be added:
\begin{align}
\mathcal{R}_{\mathrm{ent}}
=
-
\sum_{t=1}^{T}
\sum_{e=1}^{N_E}
\omega_t^{(e)}
\log \omega_t^{(e)}.\label{p36}%
\end{align}
Then the final training objective becomes
\begin{align}
\min_{\{\Theta_e\},\boldsymbol{\phi}}
\sum_{e=1}^{N_E}
\mathcal{L}_e
+
\lambda_g \mathcal{L}_{\mathrm{gate}}
-
\lambda_h \mathcal{R}_{\mathrm{ent}},\label{p37}%
\end{align}
where $\lambda_g$ and $\lambda_h$ are non-negative coefficients. This training design is more complete than simply averaging experts: each expert is encouraged to specialize, while the gating network learns to route different historical patterns to appropriate experts.

\subsection{Reliability-Aware Monitoring Decision}
After obtaining the aggregated grant probability $p_t$, the IoT-D determines whether to monitor the PDCCH through a reliability-aware threshold rule:
\begin{align}
a_t =
\begin{cases}
1, & p_t \geq \rho_t,\\
0, & p_t < \rho_t,
\end{cases}\label{p38}%
\end{align}
where $a_t=1$ means monitoring the PDCCH, while $a_t=0$ means skipping PDCCH monitoring. The threshold $\rho_t$ is dynamically adjusted according to the empirical FNR. Define
\begin{align}
\widehat{\mathrm{FNR}}_t
=
\frac{
\sum_{\tau=1}^{t} G_\tau(1-a_\tau)
}{
\sum_{\tau=1}^{t} G_\tau+\delta
},\label{p39}%
\end{align}
where $\delta>0$ avoids division by zero. The reliability controller is updated as
\begin{align}
\eta_{t+1}
=
\left[
\eta_t
+
\beta
\left(
\widehat{\mathrm{FNR}}_t-\epsilon
\right)
\right]^+ ,\label{p40}%
\end{align}
where $\epsilon$ is the target FNR threshold and $\beta$ is the update step size. The decision threshold is $\rho_t
=
\left[
\rho_0
-
c\eta_t
\right]_{\rho_{\min}}^{\rho_{\max}}$, where $c>0$ is the threshold-control coefficient, and $[x]_{\rho_{\min}}^{\rho_{\max}}
=
\min\{\rho_{\max},\max\{\rho_{\min},x\}\}$.
This bounded projection prevents the threshold from becoming too small or too large. The interpretation is direct. If the empirical FNR exceeds the required threshold, then $\eta_t$ increases, which decreases $\rho_t$. A smaller threshold makes the IoT-D more likely to monitor the PDCCH, thereby reducing missed grants. If the empirical FNR is safely below the threshold, the threshold can remain higher, allowing the IoT-D to skip more low-probability TTIs to save energy.

\subsection{Energy-Aware Refinement of the Decision Rule}

A pure probability-threshold rule may ignore the actual energy difference between monitoring and sleeping. Therefore, the threshold can be further derived from the expected one-step cost. Let the energy cost of monitoring be
\begin{align}
C_t^{\mathrm{mon}}
=
E_t^{\mathrm{op}}(a_t=1)
+
E_t^{\mathrm{sw}}(a_t=1),\label{p41}%
\end{align}
and the energy cost of sleeping be
\begin{align}
C_t^{\mathrm{slp}}
=
E_t^{\mathrm{op}}(a_t=0)
+
E_t^{\mathrm{sw}}(a_t=0).\label{p42}%
\end{align}
If the IoT-D sleeps while a grant exists, it incurs a missed-grant penalty $\lambda$. Therefore, the expected sleeping cost is
\begin{align}
\bar{C}_t^{\mathrm{slp}}
=
C_t^{\mathrm{slp}}
+
\lambda p_t
+
\eta_t p_t .\label{p43}%
\end{align}
The monitoring cost is
$\bar{C}_t^{\mathrm{mon}}
=
C_t^{\mathrm{mon}}$. The IoT-D monitors when $\bar{C}_t^{\mathrm{mon}}
\leq
\bar{C}_t^{\mathrm{slp}}$. Substituting the two costs gives $C_t^{\mathrm{mon}}
\leq
C_t^{\mathrm{slp}}
+
(\lambda+\eta_t)p_t$.
Thus, $p_t
\geq
(C_t^{\mathrm{mon}}-C_t^{\mathrm{slp}})/
(\lambda+\eta_t)$.
Therefore, an energy-aware dynamic threshold can be defined as
\begin{align}
\rho_t^{E}
=
\left[
\frac{
C_t^{\mathrm{mon}}-C_t^{\mathrm{slp}}
}{
\lambda+\eta_t
}
\right]_{\rho_{\min}}^{\rho_{\max}} .\label{p44}%
\end{align}
Then the action becomes
\begin{align}
a_t =
\begin{cases}
1, & p_t \geq \rho_t^{E},\\
0, & p_t < \rho_t^{E}.
\end{cases}\label{p45}%
\end{align}
Once the monitoring action is determined, it is mapped to the modem component states. For the components required for PDCCH blind detection, $q_{m,t}=a_t$, $m \in \mathcal{M}_{\mathrm{PDCCH}}$. For other components, $q_{m,t}
=
\varphi_m(a_t,q_{m,t-1})$,
$m \in \mathcal{M}\setminus\mathcal{M}_{\mathrm{PDCCH}}$,
where $\varphi_m(\cdot)$ denotes the component-specific DPM transition rule. 
The The Proposed MoE-IOHMM Predictive Monitoring Scheme is summarized in Algorithm~\ref{alg:moe_iohmm}. 
\begin{algorithm}[t]
\caption{The Proposed MoE-IOHMM Predictive Monitoring Scheme}
\label{alg:moe_iohmm}
\begin{algorithmic}[1]
\REQUIRE Historical IoT-D-side traces, observation window length $L$, IO-HMM order $K$, number of experts $N_E$, FNR threshold $\epsilon$, missed-grant penalty $\lambda$, reliability step size $\beta$, threshold bounds $\rho_{\min}$, $\rho_{\max}$.
\ENSURE Trained IO-HMM experts, gating network, reliability-aware threshold controller, and IoT-D-side PDCCH monitoring policy.
\STATE \textbf{Initialization:} Initialize IO-HMM expert parameters $\{\Theta_e\}_{e=1}^{N_E}$, gating parameter $\boldsymbol{\phi}$, reliability variable $\eta_0$, hidden-state beliefs $\{\alpha_0^{(e)}\}_{e=1}^{N_E}$, and modem component states.
\STATE \textbf{Offline training phase:}
\STATE Construct training samples $\{\mathcal{H}_t,G_t\}_{t=1}^{T}$ from historical traces.
\FOR{each training epoch}
    \FOR{each TTI $t$}
        \STATE Each IO-HMM expert performs one-step prediction:
        $\widetilde{\alpha}_{t}^{(e)}(j)
        =
        \sum_i
        \alpha_{t-1}^{(e)}(i)
        A_{ij}^{(e)}(\mathbf{V}_t)$.
        \STATE Each expert outputs
        $p_t^{(e)}
        =
        \sum_j
        \widetilde{\alpha}_{t}^{(e)}(j)
        B_j^{(e)}(1|\mathbf{V}_t)$.
        \STATE The gating network computes
        $\omega_t^{(e)}
        =\exp(r_e(\mathcal{H}_t;$ $\boldsymbol{\phi}))/
        (\sum_{\ell=1}^{N_E}
        \exp(r_\ell(\mathcal{H}_t;\boldsymbol{\phi})))$.
        \STATE Aggregate expert predictions:
        $p_t
        =
        \sum_{e=1}^{N_E}
        \omega_t^{(e)}p_t^{(e)}$.

        \STATE Compute expert responsibility
        $\gamma_t^{(e)}
        =
        (\omega_t^{(e)}(p_t^{(e)})^{G_t}$ $
        (1-p_t^{(e)})^{1-G_t})
        /(\sum_{\ell=1}^{N_E}
        \omega_t^{(\ell)}
       (p_t^{(\ell)})^{G_t}
        (1-p_t^{(\ell)})^{1-G_t})
        $.
    \ENDFOR

    \STATE Update IO-HMM experts and gating network by minimizing
    $\sum_{e=1}^{N_E}
    \mathcal{L}_e
    +
    \lambda_g \mathcal{L}_{\mathrm{gate}}
    -
    \lambda_h \mathcal{R}_{\mathrm{ent}}$.
\ENDFOR

\STATE \textbf{Online monitoring phase:}
\FOR{each TTI $t$}
    \STATE Construct $\mathcal{H}_t$ and $\mathbf{V}_t$. Each IO-HMM expert predicts $p_t^{(e)}$. The gating network computes $\omega_t^{(e)}$. Obtain the final grant probability
    $p_t=\sum_{e=1}^{N_E}\omega_t^{(e)}p_t^{(e)}$. Compute the energy-aware reliability threshold $\rho_t^E
    =[(C_t^{\mathrm{mon}}-$
    $C_t^{\mathrm{slp}})/
    (\lambda+\eta_t)]_{\rho_{\min}}^{\rho_{\max}}$.
    \STATE Determine the PDCCH monitoring action, if
    $a_t=1$, $p_t \geq \rho_t^E$, otherwise $a_t=1$, $p_t < \rho_t^E$.
    \STATE Map $a_t$ to component states $\{q_{m,t}\}$. Execute PDCCH monitoring or sleeping. Record the grant outcome $G_t$ and missed-grant indicator $M_t = G_t(1-a_t)$. Update empirical FNR
    $\widehat{\mathrm{FNR}}_t=(\sum_{\tau=1}^{t} G_\tau(1-a_\tau))/(\sum_{\tau=1}^{t}$ $G_\tau+\delta)$.
    \STATE Update the reliability variable
    $\eta_{t+1}
    =[\eta_t+\beta(
    \widehat{\mathrm{FNR}}_t-$ $\epsilon)]^+$.
\ENDFOR

\end{algorithmic}
\end{algorithm}

\section{Numerical Results}\label{V}
In the simulations, we evaluate the IoT-D-side PDCCH monitoring performance over a finite horizon of $T=10000$ TTIs. Four representative scheduling-grant patterns are considered, namely bursty, mixed, periodic, and sparse traffic, to emulate heterogeneous 5G NR grant-arrival behaviors. The compared policies include Always Monitor, Always Sleep, Fixed Period monitoring with $N=2$,$5$,$10$, History-Based monitoring with window sizes $5$ and $10$, Random monitoring with probabilities $p=0.2$,$0.5$,$0.8$, and the Oracle policy. The Always Monitor policy is used as the energy baseline with zero missed grants, while the Oracle policy serves as the ideal lower bound that assumes perfect non-causal knowledge of grant arrivals. The performance metrics include total energy consumption, energy-saving ratio, FNR, monitoring ratio, missed grants, and switching count. This setting is consistent with the proposed reliability-constrained PDCCH monitoring framework, where the IoT-D minimizes receiver operating energy and switching energy while controlling missed-grant events under an FNR constraint.

\begin{figure*}[htbp!]
  \centering
  \includegraphics[width=0.75\textwidth, height=0.35\textwidth]{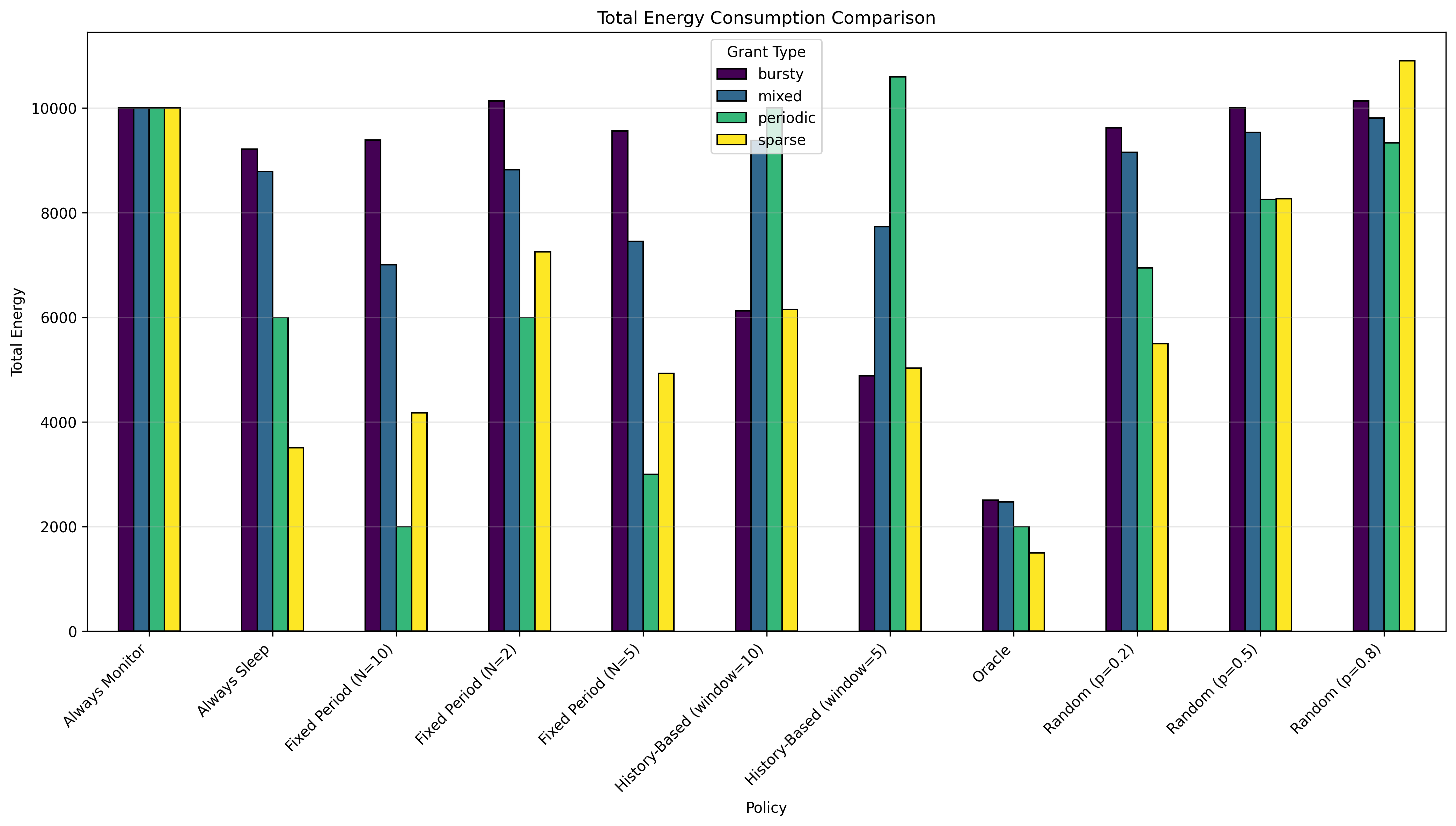}
  \captionsetup{justification=centering}
  \caption{Total Energy Consumption Comparison of Different PDCCH Monitoring Policies under Heterogeneous Grant Patterns.}\vspace{-10pt}
\label{FIGURETS2}
\end{figure*}

Fig.~\ref{FIGURETS2} compares the total energy consumption of different PDCCH monitoring policies under bursty, mixed, periodic, and sparse grant-arrival patterns. The Always Monitor policy consumes 10,000 energy units for all traffic types because the IoT-D keeps monitoring the PDCCH in every TTI, which guarantees zero FNR but provides no energy saving. In contrast, the Oracle policy achieves the lowest energy consumption, reducing the total energy to 2506.30, 2470.00, 1999.95, and 1497.90 under bursty, mixed, periodic, and sparse traffic, respectively, corresponding to energy-saving ratios of $74.9\%$, $75.3\%$, $80.0\%$, and $85.0\%$. Fixed-period policies can save energy in some regular scenarios, especially for periodic traffic, but their performance is highly sensitive to the monitoring interval and traffic pattern. For example, Fixed Period ($N=10$) achieves $80.0\%$ energy saving under periodic traffic but suffers from high FNR under bursty and sparse traffic. History-based policies exploit short-term grant correlations and reduce energy consumption under bursty and sparse traffic, but they are still unstable across different grant patterns. Random policies are generally inefficient because they ignore temporal grant structures; increasing the monitoring probability reduces missed grants but also increases energy consumption and switching overhead. These results demonstrate that simple fixed, random, or handcrafted history-based policies cannot consistently balance energy saving and reliability across heterogeneous traffic patterns, which motivates the proposed MoE-IOHMM-based predictive monitoring design.

\begin{figure}[htbp!]
  \centering
  \includegraphics[width=0.5\textwidth, height=0.35\textwidth]{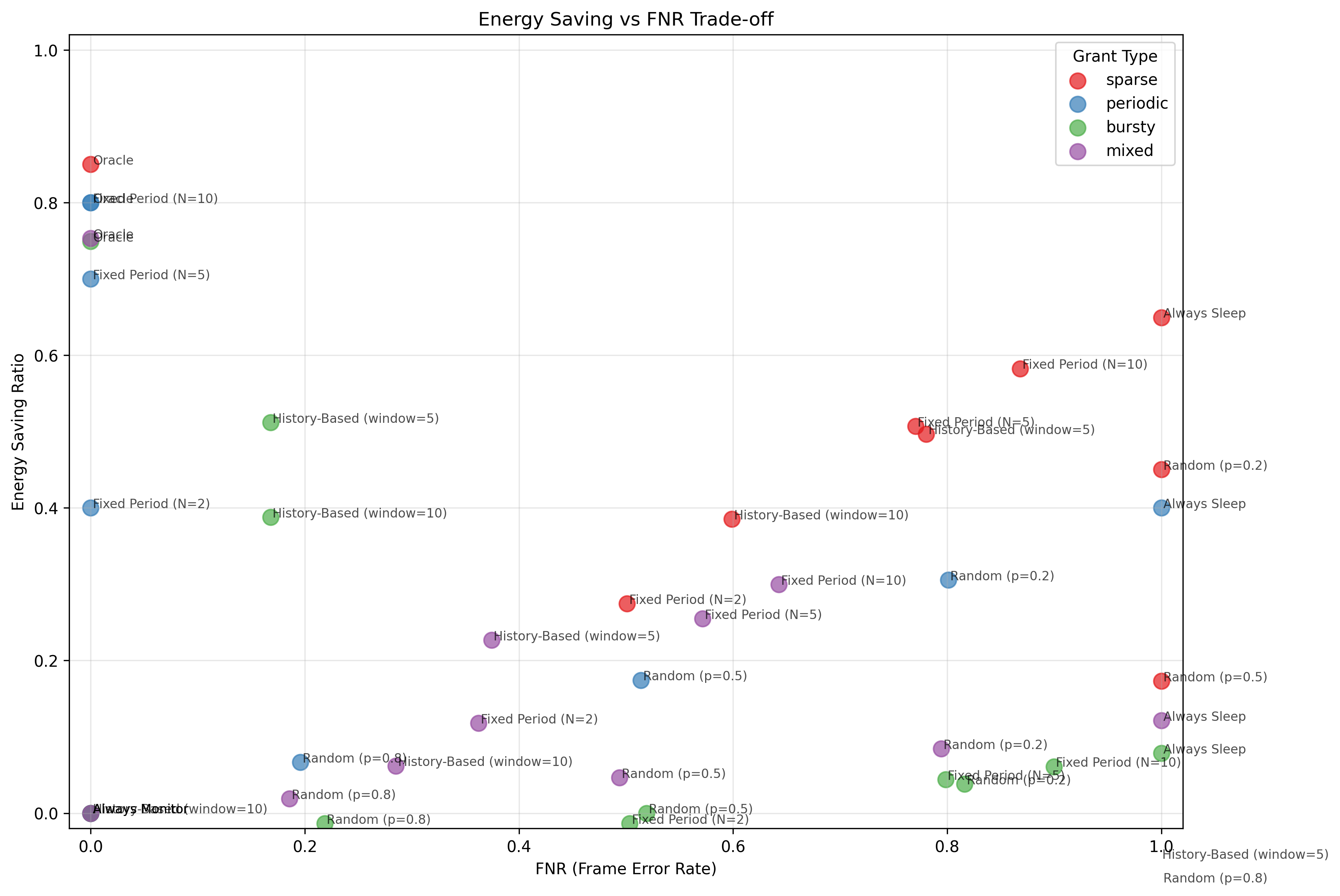}
  \captionsetup{justification=centering}
  \caption{Energy-Saving and Reliability Tradeoff of Different PDCCH Monitoring Policies under Heterogeneous Grant Patterns.}\vspace{-10pt}
\label{FIGURETS3}
\end{figure}
Fig.~\ref{FIGURETS3} further illustrates the tradeoff between the energy-saving ratio and the false negative rate (FNR) under different grant-arrival patterns. The horizontal axis denotes the FNR caused by missed scheduling grants, while the vertical axis represents the corresponding energy-saving ratio. The Always Monitor policy is located near the origin, since it monitors the PDCCH in every TTI and therefore achieves zero FNR but almost no energy saving. In contrast, the Always Sleep policy obtains a high energy-saving ratio, especially under sparse and periodic traffic, but it also leads to a very high FNR because most valid grants are missed. The Oracle policy appears in the upper-left region, achieving a high energy-saving ratio with nearly zero FNR, and thus serves as the ideal performance upper bound. Fixed-period and random policies show scattered performance across different traffic types, indicating that their energy-reliability tradeoff is highly sensitive to the grant pattern and parameter setting. For example, a larger sleeping interval or a smaller random monitoring probability can improve energy saving, but it usually causes a significant increase in FNR. History-based policies can exploit temporal grant correlations and achieve better tradeoffs in some bursty or mixed scenarios, but their performance is still inconsistent across different traffic types. These results clearly demonstrate that energy saving and grant-detection reliability are strongly coupled: aggressive sleeping improves energy efficiency but increases missed grants, whereas conservative monitoring reduces FNR at the cost of higher energy consumption. Therefore, a reliable predictive monitoring scheme should not only minimize energy consumption, but also explicitly control the FNR, which supports the necessity of the proposed reliability-aware MoE-IOHMM monitoring framework.

\begin{figure*}[htbp!]
  \centering
  \includegraphics[width=0.75\textwidth, height=0.35\textwidth]{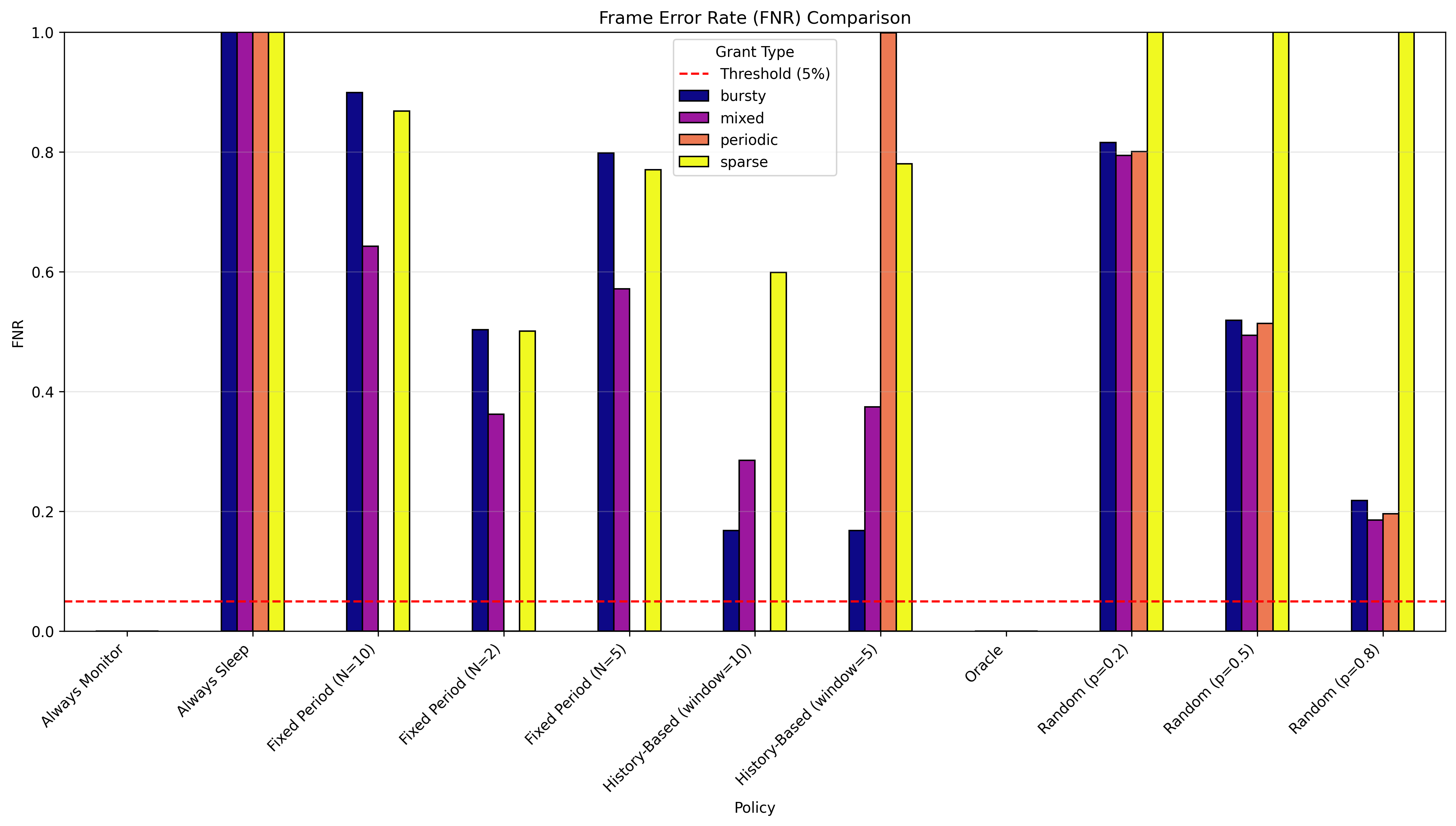}
  \captionsetup{justification=centering}
  \caption{False Negative Rate Comparison of Different PDCCH Monitoring Policies under Heterogeneous Grant Patterns.}\vspace{-10pt}
\label{FIGURETS4}
\end{figure*}
Fig.~\ref{FIGURETS4} compares the false negative rate (FNR) of different PDCCH monitoring policies under bursty, mixed, periodic, and sparse grant-arrival patterns, where the red dashed line denotes the prescribed reliability threshold of $5\%$. The Always Monitor and Oracle policies achieve zero FNR for all traffic types, indicating that no valid scheduling grants are missed. However, Always Monitor obtains this reliability by continuously activating the receiver chain, while Oracle represents an ideal non-causal lower bound with perfect grant knowledge. In contrast, Always Sleep leads to an FNR of 1 for all grant patterns, because the IoT-D skips PDCCH monitoring in every TTI and therefore misses all valid grants. Fixed-period policies can reduce FNR when the monitoring interval is short, but their reliability is still far above the $5\%$ threshold in most cases. For example, Fixed Period (N=2) reduces the FNR compared with N=5 and N=10, but still causes considerable missed grants under bursty, mixed, and sparse traffic. History-based policies achieve better reliability than fixed-period and random policies in some cases, especially when the historical window matches the temporal structure of grant arrivals. Nevertheless, their FNR remains highly traffic-dependent and often exceeds the reliability threshold. Random policies also fail to provide reliable control: increasing the monitoring probability from $p=0.2$ to $p=0.8$ reduces FNR for bursty, mixed, and periodic traffic, but the FNR under sparse traffic remains close to 1 because random monitoring cannot capture rare grant occurrences effectively. These results demonstrate that energy-saving policies based only on fixed rules, random actions, or simple historical windows cannot guarantee the required missed-grant reliability. Therefore, the FNR constraint in the proposed reliability-constrained formulation is necessary, and the monitoring decision should be guided by an adaptive grant-probability predictor and a reliability-aware threshold controller, as designed in the proposed MoE-IOHMM framework.

\begin{figure}[htbp!]
  \centering
  \includegraphics[width=0.5\textwidth, height=0.35\textwidth]{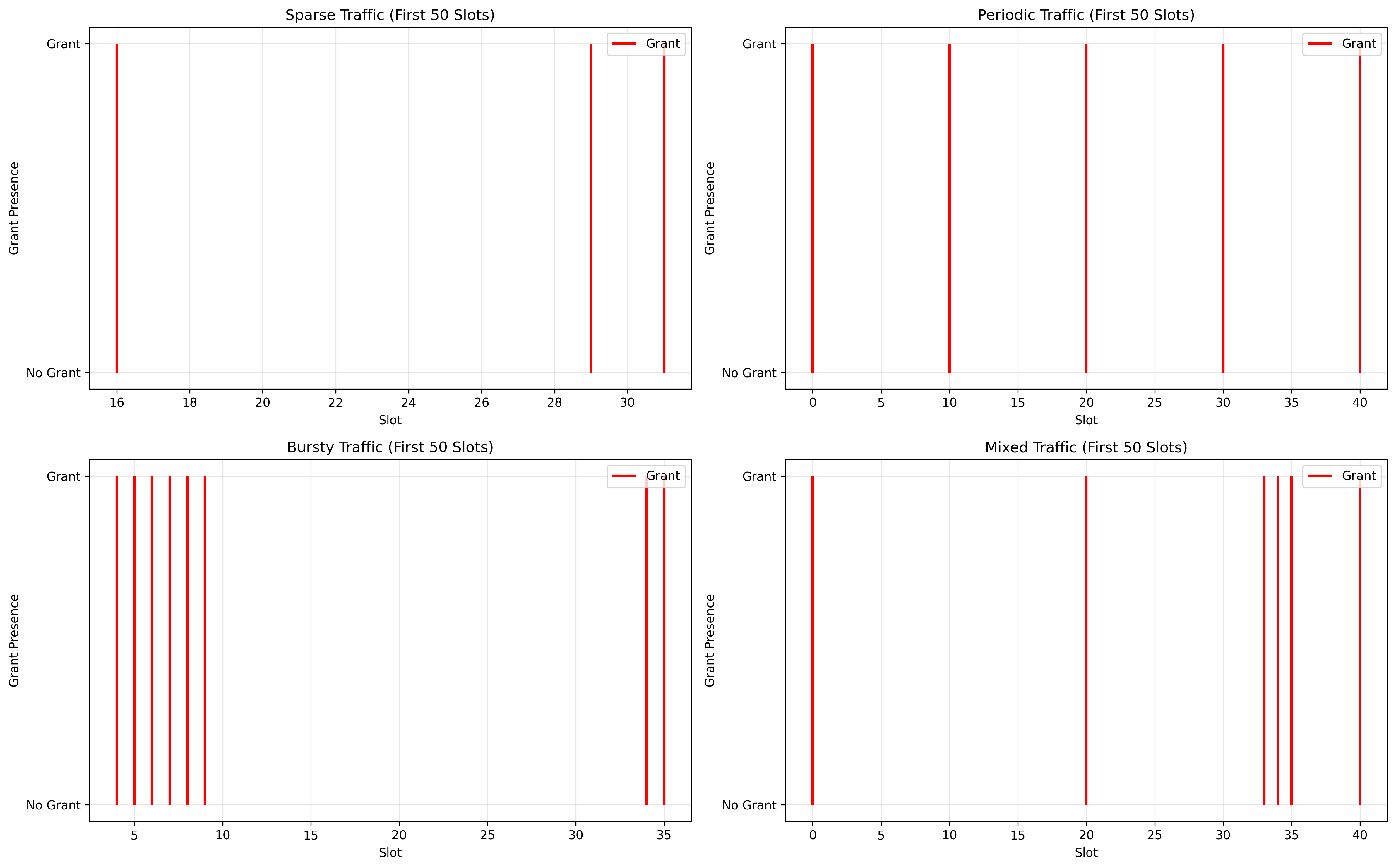}
  \captionsetup{justification=centering}
  \caption{Representative Grant-Arrival Sequences under Sparse, Periodic, Bursty, and Mixed Traffic Patterns.}\vspace{-10pt}
\label{FIGURETS5}
\end{figure}
Fig.~\ref{FIGURETS5} illustrates the first 50 slots of four representative scheduling-grant sequences, including sparse, periodic, bursty, and mixed traffic. The sparse traffic pattern contains only a few isolated grant events separated by long grant-free intervals, which reflects low-activity IoT or delay-tolerant traffic and provides large potential for IoT-D energy saving. The periodic traffic pattern shows regularly repeated grant arrivals, where deterministic monitoring policies may perform well if their monitoring period is well aligned with the grant interval. In contrast, the bursty traffic pattern exhibits consecutive grants within a short time window followed by long inactive periods, indicating strong temporal correlation and sudden scheduling activity. The mixed traffic pattern combines isolated grants, periodic-like arrivals, and short bursts, making it more difficult for fixed or random monitoring rules to achieve stable performance. These observations explain why different baseline policies show highly traffic-dependent energy and FNR performance in the previous figures. Since the IoT-D must decide whether to monitor the PDCCH before observing the actual grant state, the monitoring policy needs to exploit historical grant structures while controlling missed grants under the FNR constraint, which is consistent with the reliability-constrained predictive monitoring formulation in this paper.

\begin{figure*}[htbp!]
  \centering
  \includegraphics[width=0.75\textwidth, height=0.3\textwidth]{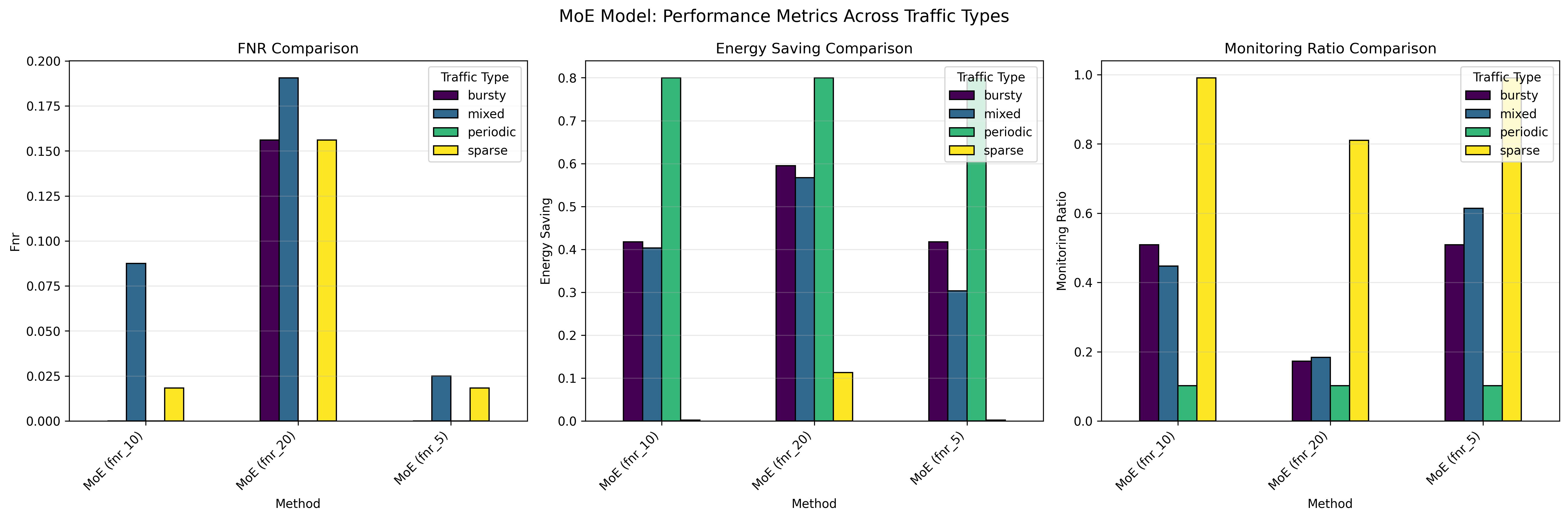}
  \captionsetup{justification=centering}
  \caption{Performance Comparison of the Proposed MoE-Based PDCCH Monitoring Scheme under Different FNR Constraints.}\vspace{-10pt}
\label{FIGURETS6}
\end{figure*}
Fig.~\ref{FIGURETS6} evaluates the performance of the proposed MoE-based PDCCH monitoring scheme under different reliability constraints, i.e., MoE with target FNR values of $10\%$, $20\%$, and $5\%$. Three key metrics are compared, including FNR, energy-saving ratio, and monitoring ratio. It can be observed that the reliability constraint directly controls the tradeoff between energy efficiency and missed-grant protection. When the FNR constraint is relaxed to $20\%$, the MoE policy becomes more aggressive in skipping PDCCH monitoring, thereby achieving higher energy-saving ratios under bursty and mixed traffic. However, this also leads to a higher FNR, especially for bursty, mixed, and sparse grant patterns. In contrast, when the FNR constraint is tightened to $5\%$, the MoE policy monitors the PDCCH more conservatively, resulting in a lower FNR but also a reduced energy-saving ratio for bursty and mixed traffic. For periodic traffic, the proposed MoE scheme consistently achieves a high energy-saving ratio of around $80\%$ with nearly zero FNR, because the periodic grant structure is easier to capture from historical observations. For sparse traffic, the monitoring ratio is close to one under strict reliability requirements, indicating that the IoT-D needs to monitor more frequently to avoid missing rare but important grant events. These results verify that the proposed MoE-based scheme can adapt its monitoring behavior according to the prescribed reliability requirement and traffic pattern, which is consistent with the reliability-aware threshold control mechanism in the proposed formulation.

\begin{figure}[htbp!]
  \centering
  \includegraphics[width=0.5\textwidth, height=0.35\textwidth]{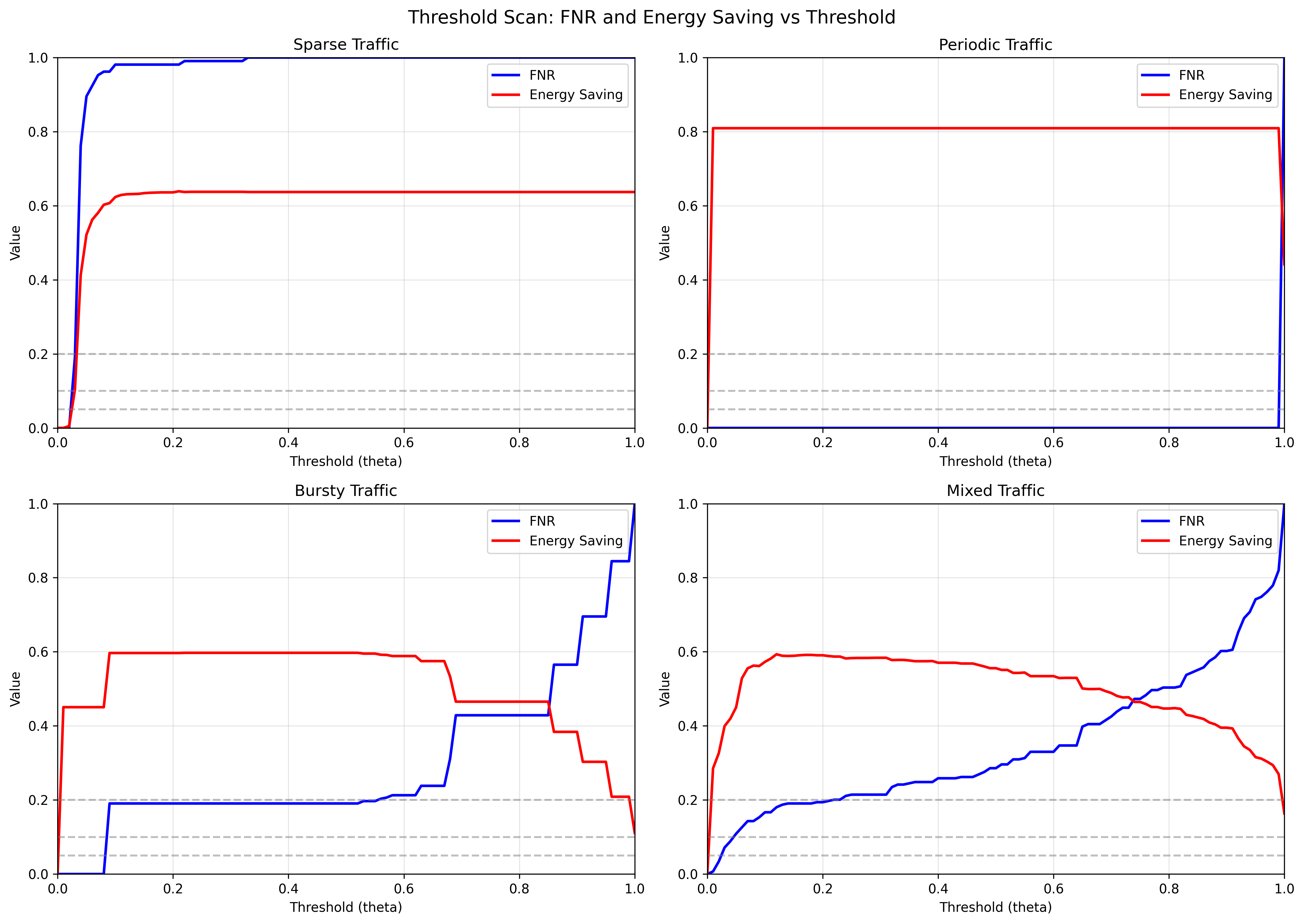}
  \captionsetup{justification=centering}
  \caption{Threshold Sensitivity of the Proposed MoE-Based Monitoring Scheme under Different Grant-Arrival Patterns.}\vspace{-10pt}
\label{FIGURETS7}
\end{figure}
Fig.~\ref{FIGURETS7} shows the threshold-scan results of the proposed MoE-based monitoring scheme, where the decision threshold $\theta$ is varied from $0$ to $1$ and the corresponding FNR and energy-saving ratio are recorded under sparse, periodic, bursty, and mixed traffic. The dashed horizontal lines indicate representative reliability levels of $5\%$, $10\%$, and $20\%$. It can be observed that different grant-arrival patterns exhibit distinct threshold sensitivities. For sparse traffic, both the FNR and energy-saving ratio increase rapidly when $\theta$ becomes slightly larger than zero, indicating that rare grant events are difficult to protect once the policy becomes aggressive. For periodic traffic, the energy-saving ratio remains high over a wide threshold range while the FNR stays close to zero, showing that periodic grants can be accurately captured from historical structures. For bursty traffic, the FNR remains relatively low under moderate thresholds but increases sharply when the threshold becomes too large, because excessive sleeping may miss consecutive grants within burst periods. For mixed traffic, the tradeoff is more gradual: increasing $\theta$ continuously improves energy saving at the cost of a steadily increasing FNR, reflecting the coexistence of isolated, periodic, and burst-like grant arrivals. These results demonstrate that a fixed threshold cannot provide a universally optimal energy-reliability balance across heterogeneous traffic patterns. Therefore, the monitoring threshold should be dynamically adjusted according to the empirical FNR and traffic characteristics, which directly supports the reliability-aware threshold controller adopted in the proposed MoE-IOHMM framework.

\begin{figure}[htbp!]
  \centering
  \includegraphics[width=0.5\textwidth, height=0.35\textwidth]{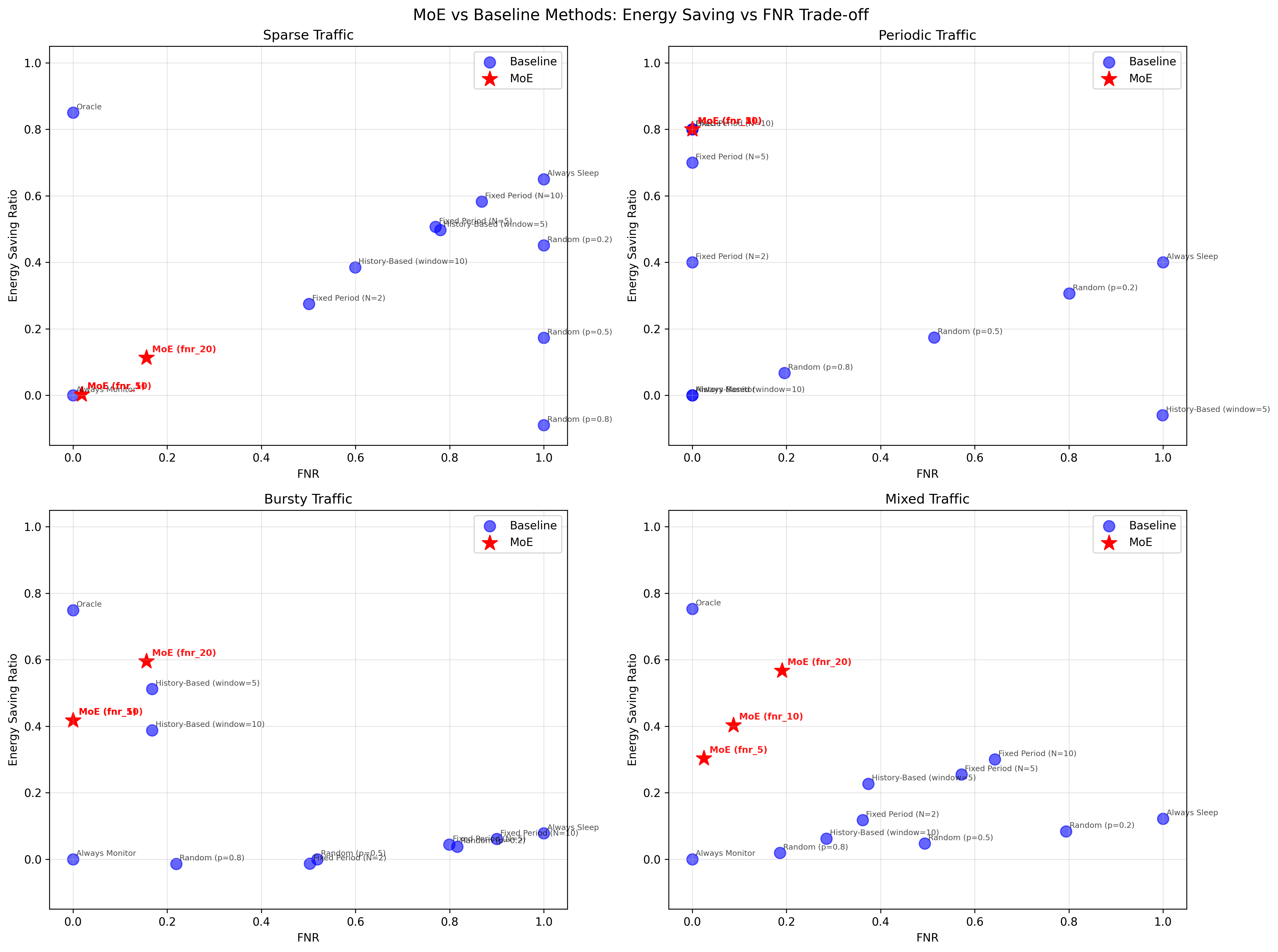}
  \captionsetup{justification=centering}
  \caption{Energy-Saving and FNR Tradeoff Comparison between the Proposed MoE Scheme and Baseline Monitoring Policies.}\vspace{-10pt}
\label{FIGURETS8}
\end{figure}
Fig.~\ref{FIGURETS8} compares the proposed MoE-based PDCCH monitoring scheme with conventional baseline policies in terms of energy-saving ratio and FNR under sparse, periodic, bursty, and mixed traffic. The blue circles denote baseline policies, including Always Monitor, Always Sleep, fixed-period monitoring, history-based monitoring, random monitoring, and Oracle, while the red stars denote the proposed MoE schemes under different target FNR constraints. It can be observed that the proposed MoE scheme achieves a more favorable energy-reliability tradeoff than most practical baselines. Under periodic traffic, the MoE schemes are located close to the upper-left region, achieving about $80\%$ energy saving with nearly zero FNR, which indicates that the proposed method can effectively learn regular grant-arrival structures. Under bursty and mixed traffic, the MoE schemes provide significantly higher energy-saving ratios than fixed-period and random policies at comparable or lower FNR levels. In particular, for mixed traffic, the MoE scheme improves the energy-saving ratio to approximately $30\%$, $40\%$, and $57\%$ under increasingly relaxed FNR constraints, whereas most baseline methods either save little energy or suffer from much larger FNR. For bursty traffic, the MoE scheme also achieves clear energy-saving gains while avoiding the extremely high-FNR region where Always Sleep and several fixed/random policies are located. For sparse traffic, strict reliability requirements force the MoE scheme to monitor more conservatively, leading to limited energy saving, which is consistent with the fact that rare grants are difficult to predict and missing them quickly increases the FNR. Overall, these results demonstrate that the proposed MoE-based monitoring policy can adapt to different grant-arrival patterns and provide a controllable tradeoff between IoT-D energy saving and missed-grant reliability, thereby validating the necessity of combining heterogeneous expert prediction with reliability-aware threshold control in the proposed framework.

\section{Conclusion}\label{VI}
In this paper, we studied the UE-side energy-saving problem for 5G NR PDCCH monitoring and formulated it as a reliability-constrained long-term energy minimization problem. By considering the fact that the IoT-D must decide whether to monitor the PDCCH before the actual grant state is observed, the proposed framework captures the essential tradeoff between receiver energy consumption and missed-grant reliability. The system model characterizes the partially observable grant-generation process using IoT-D-side historical observations, while the optimization problem jointly accounts for modem operating energy, switching energy, and false-negative-rate constraints. To address the heterogeneity of grant patterns across applications and scenarios, a MoE-IOHMM-based predictive monitoring scheme was further introduced, where multiple IO-HMM experts estimate grant probabilities and a gating network adaptively combines their outputs. The resulting probability is converted into a monitoring action through a reliability-aware threshold controller, enabling the IoT-D to reduce unnecessary receiver activation while maintaining the missed-grant ratio within an acceptable range. Overall, this framework provides an interpretable and practical solution for predictive dynamic power management in energy-efficient 5G NR terminals.


\end{document}